# Three-dimensional in situ characterization of phase transformation induced austenite grain refinement in nickel-titanium

A.N. Bucsek[a,1], L. Casalena[b,2], D.C. Pagan[c], P.P. Paul[d], Y. Chumlyakov[e], M.J. Mills[b], A.P. Stebner[a,*]

[a]Mechanical Engineering, Colorado School of Mines, 1500 Illinois St., Golden, CO 80401, USA.
[b]Materials Science & Engineering, The Ohio State University, 2041 N. College Rd., Columbus, OH 43210, USA.
[c]Cornell High Energy Synchrotron Source, 161 Wilson Laboratory, Ithaca, NY 14853, USA.
[d]Mechanical Engineering, Northwestern University, 2145 Sheridan Rd., Evanston, IL 60208, USA.
[e]Department of Physics of Metals, Tomsk State University, Lenin Ave. 36, Tomsk, Russia 634050.
*Corresponding author: astebner@mines.edu.

[1]Present address: Aerospace Engineering and Mechanics, University of Minnesota, 110 Union St. SE, Minneapolis, MN 55455, USA.
[2]Present address: Thermo Fisher Scientific, 5350 NE Dawson Creek Dr., Hillsboro, OR 97142, USA.

**Abstract:** Near-field and far-field high-energy diffraction microscopy and microcomputed tomography X-ray techniques were used to study a bulk single crystal nickel-titanium shape memory alloy sample subjected to thermal cycling under a constant applied load. Three-dimensional in situ reconstructions of the austenite microstructure are presented, including the structure and distribution of emergent grain boundaries. After one cycle, the subgrain structure is significantly refined, and heterogeneous $\Sigma 3$ and $\Sigma 9$ grain boundaries emerge. The low volume and uneven dispersion of the emergent $\Sigma$ boundaries across the volume show why previous transmission electron microscopy investigations of $\Sigma$ grain boundary formation were inconsistent.

**Keywords:** shape memory alloys, coincident lattice, grain refining, X-ray diffraction, 3D reconstruction

Actuation is an important functional ability of shape memory alloys (SMAs). Solid-state SMA actuators are compact, scalable, and capable of supporting and generating large mechanical loads while weighing a fraction of more traditional actuators. Hence, SMAs are an enabling technology for nanoelectromechanical and microelectromechanical systems (NEMS and MEMS), biomedical, active damping, and aerospace actuation systems [1–8]. The standard SMA actuation characterization procedure is constant force thermal cycling: a constant load is applied to the material, and then temperature is cycled above and below the transformation temperatures, resulting in reversible transformation strains to accumulate and recover as the material transforms between its austenite and martensite phases [9,10]. The vast majority of SMAs, however, do not exhibit perfectly reversible actuation performances. There is an inherent cyclic instability that causes the mechanical response of an SMA to change from cycle to cycle. This instability is known as functional fatigue and is understood to originate from defects such as dislocations and grain boundaries that are generated during the phase transformation process [11,12].

Bowers et al. [13] studied the microstructure evolution of actuated nickel-titanium (NiTi) SMAs with scanning transmission electron microscopy (STEM) diffraction contrast imaging and automated crystal orientation and phase mapping (ACOM)-TEM-based techniques. They found that a high density of dislocations form in the austenite phase after just a few cycles and that grain boundaries form in the austenite with large, well-defined misorientations (Σ values) based on coincident site lattice (CSL) theory after 20–100 cycles. They clearly demonstrated the possibility of these special Σ boundaries to emerge in the austenite phases; however, the small sampling sizes required for electron microscopy (a few μm) prevented them from reporting how many actuation cycles were needed for the special Σ boundaries to form, how prevalent the boundaries were, or what a representative volume is for observing these boundaries. Subsequently, Casalena [14] observed Σ3 boundaries in one section of a sample that had been cycled twice but did not observe any special Σ boundaries in another section using transmission Kikuchi diffraction (TKD). Special Σ boundaries were also not observed in other samples cycled 20 and 100 times, and only two small Σ9 boundaries were observed in a sample cycled over 20,000 times [14].

The in situ experiments presented in this letter were motivated to better understand the reasons for inconsistencies in previous experiments where emergent Σ boundaries were observed in samples with only a few cycles, but not observed at all in samples actuated tens of thousands of cycles. The process of austenite grain refinement is linked to functional fatigue, which is a primary limitation in adopting SMA actuation technologies [15–17]. Understanding how, why, and where the new Σ grain boundaries emerge is critical to engineering new materials and/or processing to limit functional fatigue. We proceed to evaluate the structure of emergent high-angle Σ and low-angle grain boundaries throughout the volume of a mm-scale single-crystal NiTi sample as a result of actuation using far-field high-energy diffraction microscopy (ff-HEDM), near-field high-energy diffraction microscopy (nf-HEDM), and microcomputed tomography (μCT).

A single crystal specimen was electrical-discharge-machined from a 40 mm diameter $Ni_{50.1}Ti_{49.9}$ single crystal ingot grown by an advanced Bridgman technique consisting of remelting a cast ingot into a graphite crucible under a helium gas atmosphere. The material exhibited a martensitic transformation from a B2 cubic austenite phase to a B19' monoclinic martensite phase. The stress-free transformation temperatures measured by differential scanning calorimetry were $M_f = -2°C$, $M_s = 15°C$, $A_s = 28°C$, and $A_f = 42°C$. The sample was machined into a rectangular dogbone tensile specimen geometry with a 1×1×1 mm³ gage section. This specimen geometry is based on the geometry used in [18] but was modified to have a 1 mm tall gage section to produce a more concentrated load in the diffracted volume [19].

The experiments were conducted on beamline F2 at the Cornell High Energy Synchrotron Source. First, the sample was heated to 120°C (above $A_f$) without a mechanical load, and the initial measurement of the austenite crystal structure (referred to as "pre-cycling") was collected. The sample was then loaded to an engineering stress of 150 MPa, cooled to 20°C, and heated back to 120°C. The second measurement of the austenite crystal structure (referred to as "after one cycle") was then collected. The mechanical load was applied using the RAMS2 load frame [20]. The heating and cooling was performed using a custom dual halogen bulb line focusing furnace with a water-cooled aluminum body and X-ray transparent upstream/downstream graphite windows. Argon was continuously flowing through the chamber during testing to minimize oxidation effects.

At the two data acquisition steps, we recorded ff-HEDM, nf-HEDM, and µCT measurements. For HEDM measurements, a volume of the sample was illuminated with monochromatic 55.618 keV X-rays while continuously rotating the sample 360º about an axis perpendicular to the incident X-ray beam direction [21–25]. Diffraction patterns were integrated in 0.25° increments, recorded on either the far-field detector or the near-field detector located 1045.7 mm and 12.91 mm from the sample, respectively. The combined nf-HEDM and ff-HEDM data can be used to spatially resolve three-dimensional (3D) orientation maps of the illuminated volume, similar to a 3D electron backscatter diffraction measurement except in situ and nondestructively. Because the orientation information in the nf-HEDM data is strongly coupled with spatial information, the reconstructions require a list of potential crystallographic orientations. The list of potential crystallographic orientations is commonly identified from the ff-HEDM data where there are fewer effects due to spatial variations. For this experiment, we also included potential orientations that were misoriented by 5° in 1° increments from each ff-HEDM orientation. This modification allows for spatial reconstructions of the intragranular misorientation [26–28]. Finally, µCT measurements of the gage section were recorded to set physical bounds for the HEDM reconstructions. The µCT technique uses X-ray absorption contrast between air and the sample to identify the edges of the sample. For all measurements, the X-ray beam was defined to 120 µm tall × 2.5 mm wide, and three data sets were recorded with the beam vertically centered at –100, 0, and +100 µm from the vertical center of the gage resulting in 1×1×0.3 mm$^3$ spatial orientation reconstructions of the gage section. The nf-HEDM orientation and spatial resolution are typically quoted as 0.1° and 2 µm, respectively [29]. (See **Supplemental Material** and corresponding Data in Brief [30] for details.)

The ff-HEDM diffraction patterns summed over all rotation increments are shown in **Fig. 1**. Three changes can be observed by comparing the patterns pre-cycling (**Fig. 1A**,**B**) and after one cycle (**Fig. 1C**,**D**). First, the reflections became more diffuse after one cycle; this is the diffraction signature of plasticity and/or grain refinement. Second, new low-intensity reflections emerged after one cycle that were not present pre-cycling. Arrows point to these new reflections in **Fig. 1C**. These new reflections are viewed more clearly by zooming into the detector regions marked by the yellow boxes (**Fig. 1B**,**D**). Note that the new reflections marked by the arrows lie along the $(110)_{B2}$ (*hkl*) ring, meaning that these reflections correspond to new austenite orientations as opposed to retained martensite. These new austenite reflections are the diffraction signatures of new grains that are highly misoriented (>15°) from the original grains. Lastly, there are low intensity B19' monoclinic reflections after one cycle (**Fig. 1D**), indicating that there is some retained martensite.

The 3D grain reconstructions shown in **Fig. 2** are spatial maps of the austenite orientation pre-cycling (**Fig. 2A**) and after one cycle (**Fig. 2B**). The orientations in **Fig. 1A**,**B** are colored according to the inverse pole figure (IPF) shown in **Fig. 2C**. The same reconstructions are shown again in **Fig. 2D**,**E** with a colormap consisting of a smaller angular range achieved by centering and stretching the IPF colormap (**Fig. 2F**). There are two significant changes in the austenite microstructure. The first change is the increase in low-angle grain boundaries, most easily seen with the high-contrast IPF colormap (compare **Fig. 2D** to **2E**). Pre-cycling (**Fig. 2D**), the grain map shows a relatively small amount of low-angle grain boundaries organized into well-structured subgrains. After one cycle (**Fig. 2E**), the subgrain structure is refined into a dense network of low-angle grain boundaries, representative of the creation and/or growth of dislocation networks that formed during the thermomechanical cycle. The second significant change in the microstructure is the emergence of new orientations that are highly misoriented from the original single-crystal orientations, most easily seen with the standard IPF colormap (compare **Fig. 2A** to **2B**). After one cycle (**Fig. 2B**), new orientations can be seen that did not exist in the original microstructure (**Fig. 2A**). These new orientations correspond to the new austenite reflections observed in **Fig. 1**. **Fig. 2G** repeats the reconstruction shown in **Fig. 2B** with the original (green) single crystal orientations made transparent to reveal the inner structure of the new grains (pink and purple). (See **Movie S1** for a full viewing of **Fig. 2G**.)

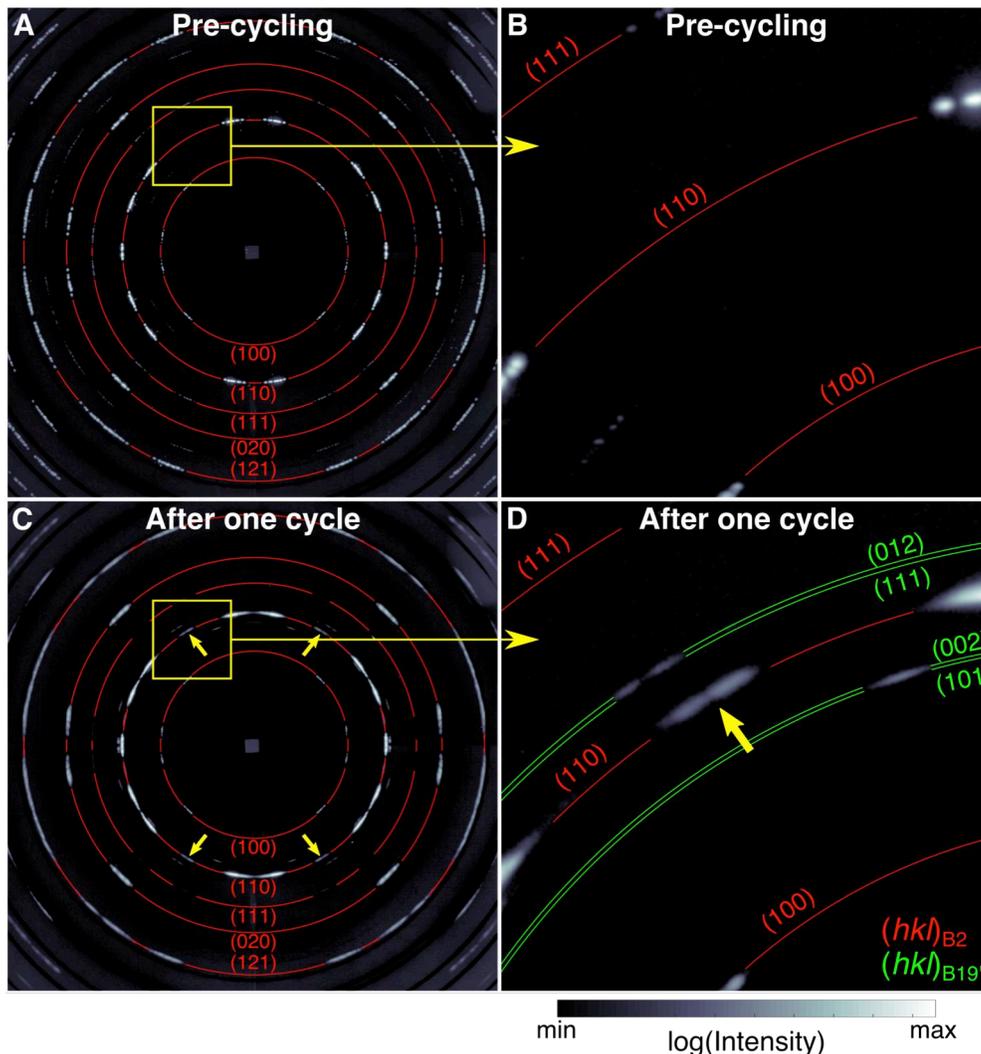

**Fig. 1.** Ff-HEDM diffraction patterns for pre-cycling (A) and after one cycle (C). The regions in the white boxes are shown zoomed-in for pre-cycling load (B) and after one cycle load (D). The B2 cubic austenite (*hkl*) rings are labeled in all subfigures, and a portion of the B19' monoclinic martensite (*hkl*) rings are labeled in (D).

The boundary misorientation angles for one 2D slice of the 3D grain reconstructions are shown in **Fig. 3**. All boundaries with misorientation angles ≥ 1° are shown, calculated as the angle of rotation needed to map the orientation of one voxel onto the neighboring voxel. Pre-cycling (**Fig. 3A**), the microstructure is near-single-crystal with organized, low-angle subgrains. The maximum subgrain angle is 13° and there is relatively little intragranular misorientation. After one cycle (**Fig. 3B**), the microstructure consists of profuse low-angle (< 5°) boundaries indicative of dense dislocation networks. The new grains (pink and purple in **Fig. 2G**) have boundary misorientation angles of roughly 59° and 39°. The distributions of boundary misorientation angles over the complete 1x1x0.3 mm$^3$ illuminated volume are shown pre-cycling (**Fig. 4A**) and after one cycle (**Fig. 4B**). A comparison of the two distributions confirms that the cycling replaced most of the 5–15° subgrain boundaries with a higher percentage of low-angle (< 5°) boundaries. The insets show the spread and relative volume of the new high-angle grain boundaries. **Table 1** shows the average misorientation angles and misorientation axes for the two new high-angle grains, calculated from the mean of the misorientation angle distributions shown in **Fig. 4** and the corresponding misorientation axes (calculated via the Rodrigues rotation formula). A comparison of the experimental boundaries and the theoretical Σ3 and Σ9 boundaries confirms that the new high-angle grains have Σ3 (pink in **Fig. 2G**, **Fig. 3B**) and Σ9 (purple in **Fig. 2G**, **Fig. 3B**) boundaries.

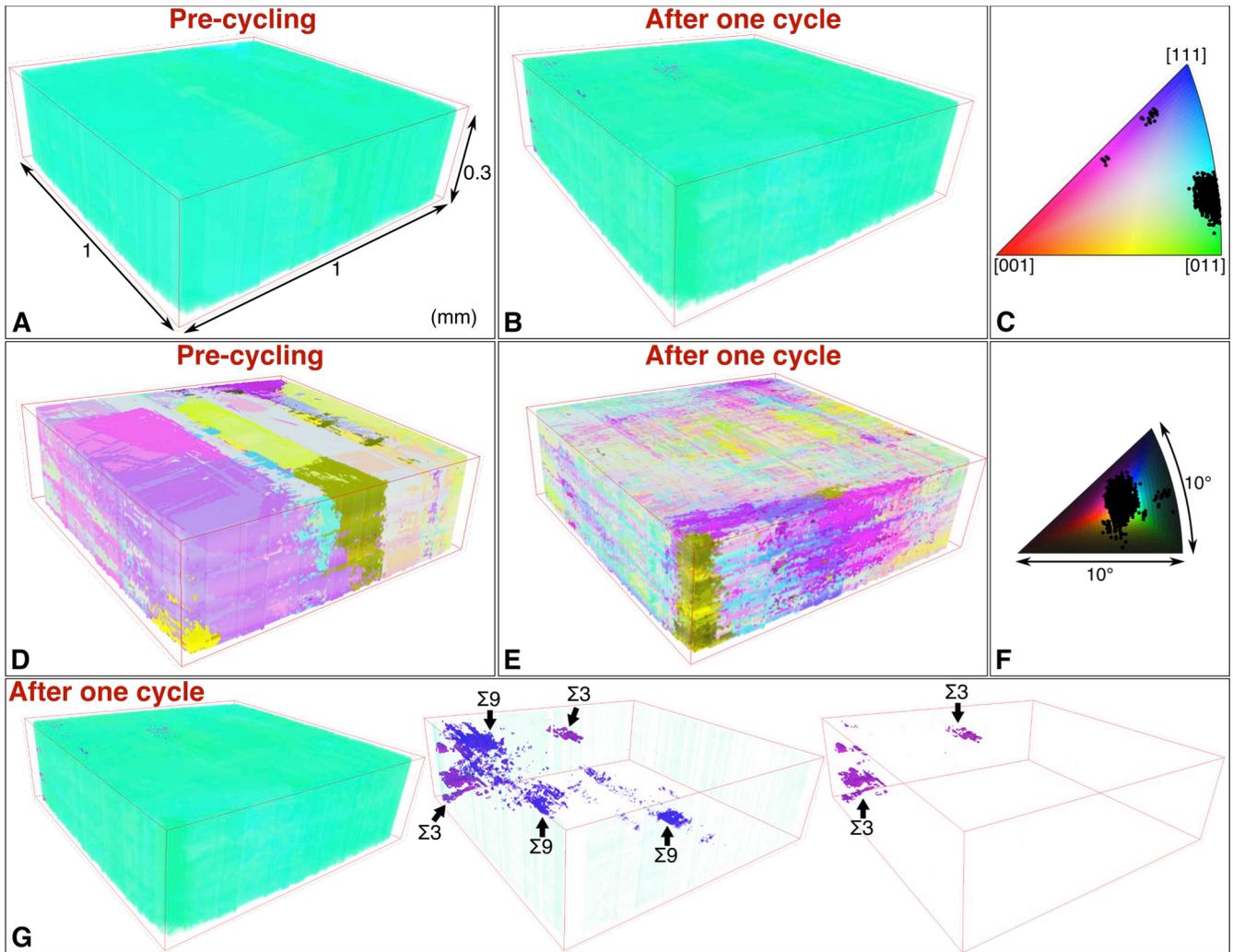

**Fig. 2.** 3D in situ grain reconstructions of a 1x1x0.3 mm$^3$ volume centered at the gage center. The colors of the pre-cycling microstructure (A) and after one cycle microstructure (B) represent the orientation of the voxel as colored by the IPF in (C). The pre-cycling (D) and after one cycle (E) grain maps are shown again with a higher angular contrast according to the centered and stretched IPF in (F). The after one cycle grain map shown in (B) is shown again in (G) with different transparencies to reveal the internal structure of the new grains. A voxel size of 5x5x5 μm$^3$ was used for all reconstructions.

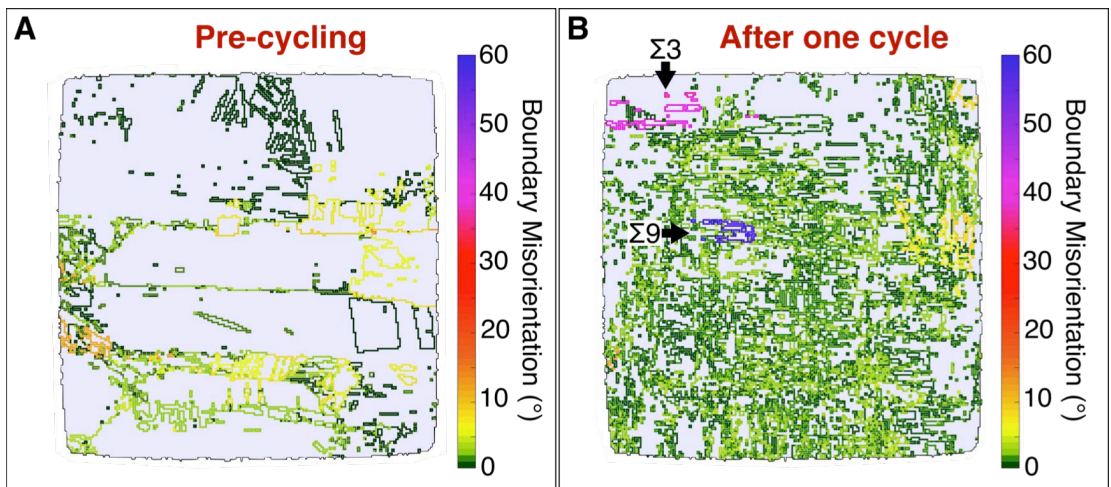

**Fig. 3.** Boundary misorientation angles for two 2D slices of the pre-cycling (A) and after one cycle (B) 3D grain reconstructions.

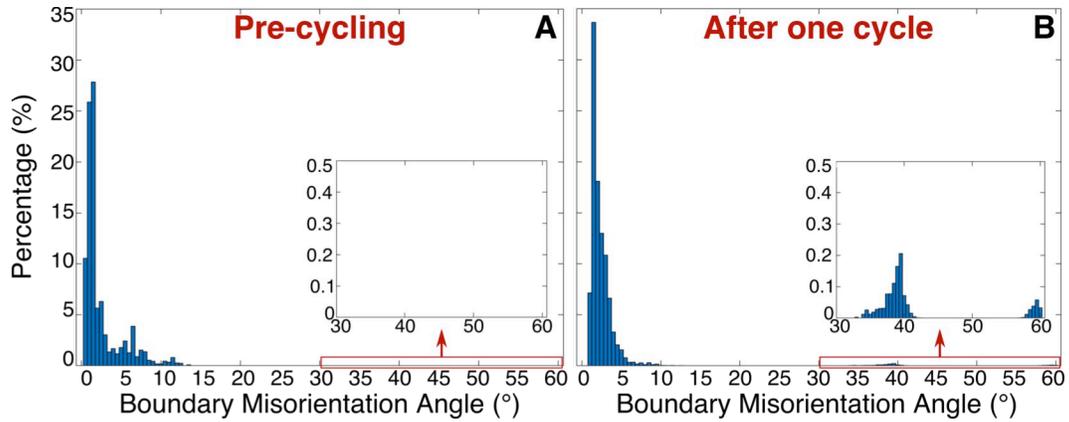

**Fig. 4.** Distributions of the boundary misorientation angles over the 3D volumes shown in **Fig. 3** pre-cycling (A) and after one cycle (B) with insets showing key high angle values with the percentage y-axis scaled from 0–0.5%.

**Table 1.** Average experimentally observed boundary misorientation axes and angles for the new grains (pink and purple in **Fig. 2B**,**G** and **Fig. 3B**), and theoretical boundary misorientation axes and angles for Σ3 and Σ9 boundaries.

| Σ | Experimental misorientation angle (°) | Theoretical misorientation angle (°) | Experimental misorientation axis | Theoretical misorientation axis |
|---|---|---|---|---|
| 3 | 59.2 | 60 | [$\overline{0.56}$ 0.56 0.61] | ⟨0.58 0.58 0.58⟩ |
| 9 | 39.3 | 38.9 | [$\overline{0.02}$ 0.69 0.72] | ⟨0 0.71 0.71⟩ |

These results show that a measurable amount of new Σ grain boundaries can form in an austenite crystal as a result of just one actuation cycle at moderate stresses and temperatures, but the emergence of these boundaries are relatively rare events during actuation under moderate loads. The low volume and uneven dispersion of the new Σ boundaries demonstrate why the observations of Σ boundary emergence were inconsistent using TEM. As can be seen in **Fig. 2**, there are relatively large volumes (hundreds by hundreds of μm) where no new Σ grains formed—a TEM sample taken from one of these large areas would have missed the new Σ grains. While these initial experimental data are not conclusive in this regard, it is probable that the stress concentrations at material imperfections such as inclusions and grain boundaries can lead to deformation twinning that produces Σ grain boundaries. This idea is similar to the observation of stress concentrations around material imperfections leading to non-optimal martensite microstructure to form [19,31]. For example, a comparison of **Fig. 3A** and **Fig. 3B** shows that the Σ3 grains formed on a subgrain boundary, and the Σ9 grains formed near some intragranular misorientation that could correspond to a defect (likely an inclusion). Still, there are many other regions that contain subgrain boundaries and misorientation that did not form Σ grains. The energetics behind these processes have started to be discussed (e.g., in [32–34]). The full 3D spatial map of orientations are provided in a corresponding Data in Brief [30]. Presumably, the formation of Σ grain boundaries will become more prevalent with higher actuation stresses and temperatures, but this hypothesis also merits further investigation. Regardless, these results suggest that minimizing heterogeneous stress concentrations within microstructures would limit the formation of new Σ grains and the functional fatigue that results from them.

Thus, it is still not clear how and when these special Σ boundaries form during actuation events, or how prevalent they are within a bulk material after actuation. It is known, however, that these special Σ boundaries correlate with deformation twin structures of both the B2 and B19′ phases. Nishida et al. [35] showed that the Σ9 boundary in the austenite phase in NiTi is the twin boundary for $\{114\}_{B2}$ and $\{201\}_{B19'}$ deformation twins, and the Σ3 B2 boundary is the twin boundary for $\{112\}_{B2}$, which corresponds with $\{113\}_{B19'}$ deformation twins [36]. Most observations of these boundaries [17] have been made after severely deforming martensite and then transforming back to austenite, though severe plastic deformation studies in the austenite state have also showed evidence of these boundaries [37]. During actuation events, it is most likely that the deformation twins form in the B19′ phase and are then inherited by the B2 phase, as twinning is much more energetically favorable in the martensite. This mechanism was experimentally confirmed by Ii et al. [38] and computationally supported using energy calculations by Ezaz et al. [32].

Recently, Wang et al. [13,33,34] proposed an alternate hypothesis that predicts that the B19′ phase can undergo shear transformations through an intermediate BCO (B33) phase that is metastable but nearby in energy state at finite temperature. During this experiment, we also attempted to identify any presence of an intermediate BCO phase to either refute or support the metastable BCO pathway hypothesized by Wang et al. [13,33,34]. The results could not prove or disprove the existence of this newly proposed mechanism and motivate the developments of future higher resolution, faster in situ experiments focused specifically on observing the parts of reciprocal space where the BCO and B19′ structures are unique. Still, all results presented in this paper are consistent with the idea that the martensite deformation twinned at local stress concentrations caused by imperfections, and those boundaries where then inherited by the B2 upon reverse transformation.

This study demonstrates the advantages of using bulk X-ray diffraction techniques such as HEDM to search for small changes in microstructure across large volumes during operating conditions. With some exceptions, HEDM has mostly been performed on high-symmetry materials with large (> 20 μm), uniform grain sizes. While previous studies illustrate how HEDM can be a useful technique to study low-symmetry materials and materials with small grain sizes [19,28,31,39], this study illustrates how HEDM can be used to investigate microstructures with significant grain size disparity. The grain size disparity required an innovative statistical approach to analyze the nf-HEDM data. The details are briefly discussed in the **Supplemental Material** and more completely described in a corresponding Data in Brief [30]. These kinds of novel approaches to HEDM data analysis procedures can expand the use of bulk 3D X-ray diffraction techniques to gain insight into a variety of material behaviors.

**Acknowledgements**

We thank Dr. Jeremy Schaffer at Fort Wayne Metals for providing the NiTi feedstock materials that were used in the Bridgman growth.

**Funding Sources**

ANB acknowledges the support provided by the National Science Foundation Graduate Research Fellowship Program (award no. DGE-1057607). APS acknowledges support provided by the National Science Foundation (award no. CMMI-1454668). ANB and APS acknowledge XSEDE resources (award no. TG-MSS160032 and TG-MSS170002). LC and MJM acknowledge support provided by the U.S. Department of Energy Office of Science (award no. DE-SC0001258). The MatCI (NSF DMR-1121262) facility at Northwestern University was used for calorimetry. This work is based upon research conducted at the Cornell High Energy Synchrotron Source (CHESS), which is supported by the National Science Foundation under awards DMR-1332208 and DMR-0936384.

**Supplementary material**

For all nf- and ff-HEDM measurements, the monochromatic 55.618 keV X-ray beam was defined to 120 μm tall × 2.5 mm wide using mechanical slits. For each of the two load steps, X-ray data were recorded with the beam vertically centered at –100, 0, and +100 μm from the vertical center of the gage. The detector data was binned at 0.25° increments for a full 360° sample rotation, resulting in 1,440 detector images for each data collection. The ff-HEDM detector, a GE41RT amorphous silicon area detector with 2048×2048 pixels and 200×200 μm² pixel size, was positioned 1.047 m from the sample. The ff-HEDM detector calibration parameters were determined using powder diffraction patterns taken at $\omega = 0°$ and $\omega = 180°$ of a NIST standard $CeO_2$ powder sample (NIST RSM 674b), where ω is the angle about which the sample is rotated as shown in Fig. 13 of [19]. The ff-HEDM detector calibration parameters were calculated using the HEDM analysis suite HEXRD [40] and are shown in **Table S1**. The nf-HEDM detector, a Retiga 4000DC camera with 2048×2048 pixels, a 1.48×1.48 μm$^2$ pixel size, and a LuAG:Ce scintillator, was positioned 12.91 mm from the sample. The near-field detector was characterized using an assembly of two offset 25×25×50 μm$^3$ gold crystals, where each crystal contained several crystal orientations. The nf-HEDM detector calibration parameters were calculated using the HEDM analysis suite HEXRD [40] and are shown in **Table S2**. Note: the near-field detector is not necessarily perfectly aligned (i.e., minor tilts with respect to the laboratory coordinate system), but we found that including sub-degree tilts of the detector in the geometric model did not influence the reconstruction likely due to the close proximity of the near-field detector to the sample. To improve the reconstructions of the sample edges, μCT measurements were taken just before each nf-HEDM measurement by removing the beam stop and taking 360 one-second exposures during the full sample rotation (1 exposure per degree of rotation). A "bright field" image (i.e., a one-second exposure without the sample in the beam) was used to normalize the μCT data for background. The μCT analysis was performed using the Inverse Radon Transform in the SciPy library [41].

The following forward model procedure was used within the HEDM analysis software HEXRD for reconstructing 3D grain maps.

1. A 3D virtual mesh of the sample is made. In this case, we used a 5×5×5 μm$^3$ voxel size.
2. For each voxel, each orientation in a list of grain orientations is forward modeled onto a virtual detector. The list of grain orientations used for the reconstruction procedure is typically the orientations indexed from a corresponding ff-HEDM data analysis. For this experiment, we also included orientations that were misoriented by 5° in 1° increments from each grain orientation indexed from the ff-HEDM analyses. This modification allows for spatial reconstructions of the intragranular misorientation [26–28].
3. The virtual detector data is compared to the experimental nf-HEDM detector data. For each sample voxel, the orientation with the highest completeness is assigned to that voxel, where completeness is defined as the percentage of virtual Bragg reflections that were verified against the experimental data. To determine which voxels of the measured volume were reliably measured, 5,000 random orientations that were known to be more than 10° misoriented from the identified grain orientations were also evaluated using the forward-model analysis methodology. The highest completeness of these testing (noise) data was 56%. Hence, a completeness threshold of 56% was used to filter all analyzed data. Voxels identified with more than 56% completeness were used in the ensuing report.

The 3D visualizations were constructed using ORS Dragonfly 2.0 [42]. The boundary misorientation analyses were performed using MTEX [43].

One important byproduct of the nf-HEDM grain reconstructions when studying microstructures with a large grain size disparity is that large grains tend to dominate small grains. When a grain is reconstructed in the virtual sample mesh, the completeness of this grain will decay after the grain boundary, schematized in one dimension in **Fig. S1**. For samples with a uniform grain size distribution, this has little effect; the completeness values will decay at comparable rates, and the grain boundary will be reconstructed accurately at the intersection (**Fig. S1B**). For samples with a bimodal grain size distribution, the completeness of very large grains can extend over neighboring small grains (**Fig. S1A**). In these cases, the large grains will be reconstructed, but the small grains will not. Furthermore, because small grains have low volume and thus low intensity, the reflections of the small grains are sometimes only detected on the brightest (*hkl*) rings, resulting in lower completeness values for the small grains than large grains and increasing the dominance of large grains over small. To capture the emergence of new, low-volume regions, we reconstructed any new orientations detected during the ff-HEDM analysis separately and then added them to the final reconstruction. In other words, we identified the low-volume regions and gave them preference over the large-volume regions in the reconstruction.

The procedure behind this approach is discussed more thoroughly in a corresponding Data in Brief [30], where the reconstructed dataset of the full 3D spatial map of orientation is also provided.

**Table S.1.** ff-HEDM detector calibration parameter values.

| Detector Calibration Parameter Name | Detector Calibration Parameter Value |
| --- | --- |
| $x$ position of beam center | –1.24 mm from detector center |
| $y$ position of beam center | 1.53 mm from detector center |
| sample-to-detector distance | 1045.7 mm |
| tilt of detector about $x$-axis | 0.064 radian |
| tilt of detector about $y$-axis | 0.052 radian |

**Table S.2** nf-HEDM detector calibration parameter values.

| Detector Calibration Parameter Name | Detector Calibration Parameter Value |
| --- | --- |
| $x$ position of beam center | 1.00 mm from detector center |
| $y$ position of beam center | –3.20 mm from detector center |
| sample-to-detector distance | 12.91 mm |
| tilt of detector about $x$-axis | 0.00 radian |
| tilt of detector about $y$-axis | 0.00 radian |

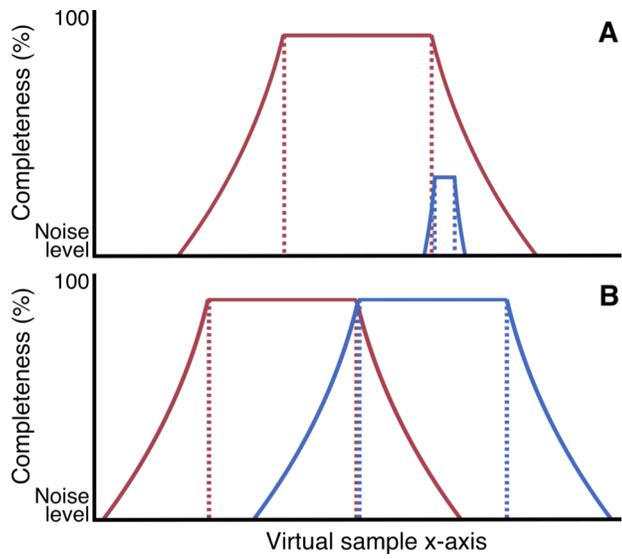
**Fig. S1.** Schematic showing the effect of size on reconstructions for disparate grain sizes (A) and similar grain sizes (B) in one dimension. The dotted lines illustrate where the grain actually ends, and the solid lines show the completeness values.